# Prediction of Turbulent Shear Stresses through Dysfunctional Bileaflet Mechanical Heart Valves using Computational Fluid Dynamics

**Fardin Khalili[\*], Peshala P.T. Gamage, Hansen A. Mansy**

[1]University of Central Florida, Department of Mechanical and Aerospace Engineering, Biomedical Acoustics Research Lab, Orlando, FL, 32816, USA

## ABSTRACT

There are more than 300,000 heart valves implanted annually worldwide with about 50% of them being mechanical valves. The heart valve replacement is often a common treatment for severe valvular disease. However, valves may dysfunction leading to adverse hemodynamic conditions. The current computational study investigated the flow around a bileaflet mechanical heart valve at different leaflet dysfunction levels of 0%, 50%, and 100%, and documented the relevant flow characteristics such as vortical structures and turbulent shear stresses. Studying the flow characteristics through these valves during their normal operation and dysfunction can lead to better understanding of their performance, possibly improved designs, and help identify conditions that may increase the potential risk of blood cell damage. Results suggested that maximum flow velocities increased with dysfunction from 2.05 to 4.49 $ms^{-1}$ which were accompanied by growing eddies and velocity fluctuations. These fluctuations led to higher turbulent shear stresses from 90 to 800 $N.m^{-2}$ as dysfunctionality increased. These stress values exceeded the thresholds corresponding to elevated risk of hemolysis and platelet activation. The regions of elevated stresses were concentrated around and downstream of the functional leaflet where high jet velocity and stronger helical structures existed.

## 1. INTRODUCTION

There are more than five million patients diagnosed with heart valve disease and about 85,000 heart valve replacements in the US annually [1,2] with half of these valves being mechanical [3]. The most common type is the bileaflet mechanical heart valve (BMHV) that has been designed, implanted [4] and gone through several design enhancements [5]. In vitro and in vivo experimental studies of BMHVs suggested that they have desirable hemodynamics and resistance to wear [6,7]. However, mechanical valves have some post-surgical complications such as hemolysis, and platelet activation [8]. They are also subject to insidious prosthetic valve dysfunction [9,10]. Failure of a mechanical heart valve can take the form of leaflets' motion restriction [11], calcification, or the appearance of cavitation [12]. Since prompt detection of insidious prosthetic valve dysfunction may help lowering mortality rates, the analysis of flow dynamics around heart valves [13–16], and cardiac sounds [17–21] have become of interest in recent studies.

Analysis of blood flow around BMHV may help identify flow patterns with higher shear stresses that can affect thrombogenic potential [22], which may therefore help improve mechanical heart valve designs. Previous studies showed that the structural failure of mechanical heart valves is usually related to blood clotting, thrombus formation and tissue overgrowth [8,23]. These complications, in overall, mostly occurred at the valve ring and hinge area and impaired the movement of leaflets [24], which in turn, may lead to a life-threatening dysfunction of one or both leaflets of BMHVs [25]. Fortunately, many of these complications can be detected with periodic monitoring of valve function [9]. Montorsi et al. [26] found that for 35% of their subjects' fluoroscopy indicated significant restriction in one of the leaflets although these patients had normal Doppler echo study. In their study, the valves with blocked leaflets could be fully recovered when valve thrombosis is detected before 21 days of thrombus occurrence. The peak velocity and the position of the maximal velocity at valve orifices were also determined as the best predictors of dysfunctional valve leaflets [27]. The practical limitations of the measurement methods in humans often make it impractical to extract detailed information and perform parametric studies. On the other hand, computational fluid dynamics (CFD) offers a complementary approach that provides a wealth of information. Studies of blood flow characteristics through a BMHV provided useful information about velocity, vortex formation, and turbulent stresses, especially around the valve hinge regions [28]. These analyses can help identify conditions that may increase the risk of blood cell damage [22]. Lethal damages of red cells can occur with turbulent shear stresses as low as 150 N.m$^{-2}$ [29], and in the presence of foreign surfaces such as valve prostheses, the threshold may be lowered significantly to 1-10 N.m$^{-2}$ [30,31]. The most commonly reported threshold for the critical turbulent shear stress is, however, 400 N.m$^{-2}$ [32].

*Corresponding Author: fardin@knights.ucf.edu

Three-dimensional shear stress analysis requires the computation of the full Reynolds stress tensor ($R$):

$$R = \begin{bmatrix} \sigma_{xx} & \tau_{xy} & \tau_{xz} \\ \tau_{yx} & \sigma_{yy} & \tau_{yz} \\ \tau_{zx} & \tau_{zy} & \sigma_{zz} \end{bmatrix} = \rho \begin{bmatrix} \overline{uu} & \overline{uv} & \overline{uw} \\ \overline{vu} & \overline{vv} & \overline{vw} \\ \overline{wu} & \overline{wv} & \overline{ww} \end{bmatrix} \qquad (1)$$

Where, $\bar{u}$, $\bar{v}$, and $\bar{w}$ are the mean velocity fluctuation components, $\rho$ is density and, $\sigma$ and $\tau$ represent normal and shear stresses, respectively. Popov [33] provides a detailed discussion of the calculation of three-dimensional maximum or principal stresses which involves the solution of the roots of the following third order equation:

$$\sigma^3 - I_1\sigma^2 + I_2\sigma - I_3 = 0 \qquad (2)$$

where,

$$I_1 = \sigma_{xx} + \sigma_{yy} + \sigma_{zz} \qquad (3)$$

$$I_2 = \sigma_{xx}\sigma_{yy} + \sigma_{yy}\sigma_{zz} + \sigma_{xx}\sigma_{zz} - \tau_{xy}{}^2 - \tau_{yz}{}^2 - \tau_{xz}{}^2 \qquad (4)$$

$$I_3 = \sigma_{xx}\sigma_{yy}\sigma_{zz} + 2\tau_{xy}\tau_{yz}\tau_{zx} - \sigma_{xx}\tau_{yz}{}^2 - \sigma_{yy}\tau_{xz}{}^2 - \sigma_{zz}\tau_{xy}{}^2 \qquad (5)$$

The three roots $\sigma_1 < \sigma_2 < \sigma_3$ of the above equation are the three principal normal stresses. The coefficients $I_1, I_2$, and , $I_3$ are functions of the measured Reynolds stress tensor and are the three stress invariants of the Reynolds stress tensor. In addition, the maximum shear stresses ($\tau_{ij_P}$) are linearly related to the normal stresses by the following equations:

$$\tau_{ij_P} = \frac{\sigma_i - \sigma_j}{2}; \ \tau_{max} = \frac{\sigma_3 - \sigma_1}{2} \qquad (6)$$

## 2. MODELS AND METHODS

### 2.1 CFD Modelling

In this study, the computational domain consists of four regions in the flow direction sequentially: upstream, heart valve, aortic root sinuses and downstream. The heart valve geometry (Fig.1a) was chosen to be similar to previous studies [1,6,34–36] to facilitate comparisons with these studies. The heart valve shown has two leaflets that, in the open position, divide the flow area into three orifices: two of these orifices (top and bottom orifices) are roughly semicircular and the third (middle orifice) is approximately rectangular, Fig. 1a. In addition, the epitrochoid shape of the asymmetric aortic sinuses geometry (from the cross-sectional view) was generated using information extracted from angiograms [37]. The entrance diameter (d) of the sinuses is 0.023 m. Creating a realistic geometry of the aortic sinuses is important for appropriate internal flow field analysis [13,38].

The CFD analysis was performed for a pulsatile flow through a three-dimensional BMHV. The inlet velocity corresponded to cardiac output of 5 L.min$^{-1}$ and heart rate of 70 bpm with a systolic phase duration of 0.3 s (Fig.1b), similar to previous studies [39,40]. The peak inflow velocity was about 1.2 ms$^{-1}$. The density and dynamic viscosity of blood were set to $\rho = 1080$ kg.m$^{-3}$ and $\mu = 0.0035$ Pa.s, respectively. These conditions correspond to an inlet peak Reynolds number ($Re_{peak} = \frac{\rho U_{peak} d_{inlet}}{\mu}$) of 8516 and a Womersley number ($W_o = \frac{d}{2}\sqrt{\omega\rho/\mu} = 26.5$; where, $\omega = \frac{2\pi}{T} = 17.21$ rad.s$^{-1}$, is the frequency of pulsatile flow and T= 0.866 s is the period. Also the outlet was set to P = 0. The Wilcox's standard-Reynolds k-Omega turbulence model [34,41–44], which is known to perform well for the flow around the bluff bodies [45], was used to simulate the flow during a complete cardiac cycle. The unsteady simulation was performed with a time step of 0.5 ms. Numerical solution typically converged to residuals about $< 10^{-4}$. Fig.1c shows the side cross section of the BMHV at different levels of leaflet dysfunction of 0, 50, and 100% (corresponding to a gradually decreasing effective orifice area) as well as the flow direction. In the current study, the simulations were run for one the entire cycle; however, the results were focused on flow characteristics around the valve at the peak systole when the two leaflets were supposed to be fully opened. The top (functional) leaflet was in the fully open position while the bottom leaflet had inability to fully open at this phase due to the leaflet immobility. Therefore, the dynamics of the leaflet opening and closure were not considered and the data presented in the results section will focus on the fully opening period from 60 to 250 ms [46].

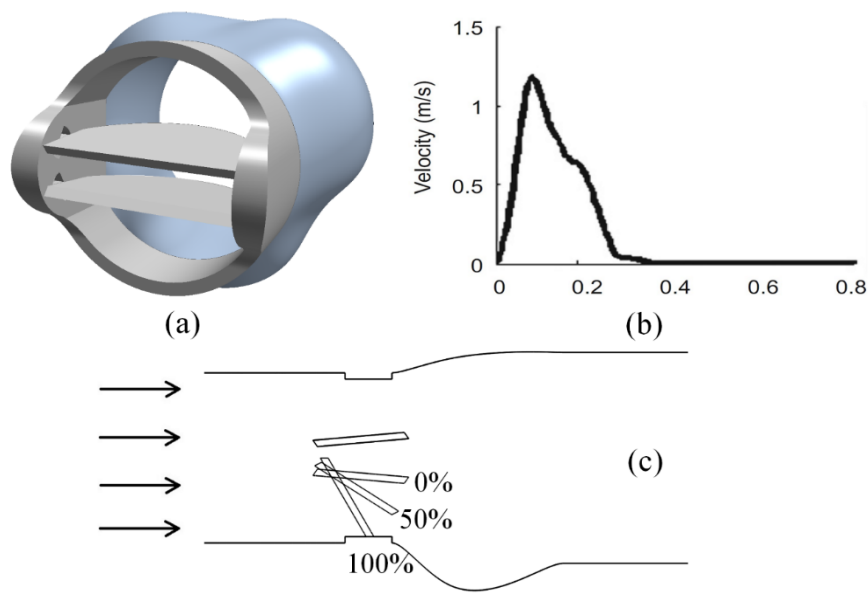

**Fig. 1** (a) Geometries of the bileaflet mechanical heart valve and aortic root sinuses; (b) Inlet velocity profile; (c) cross-sectional view of flow domain and the valve leaflets with different levels of dysfunction of the bottom leaflet

## 2.2 Mesh Generation and Validation

Three mesh configurations were considered to find the mesh-independent solution for this study (Fig. 2a). Mesh 1, 2, and 3 had the numbers of cells of approximately 800,000, 1.2 million, and 3 million. For this analysis, the velocity profile was considered at the entrance of the aortic sinuses, as shown in this figure. The similarity in the results for mesh 2 and mesh 3 showed that the velocity profile did not change. This high quality polyhedral mesh was refined especially in the heart valve and aortic sinuses regions. In addition, to maintain $y+$ less than 1 close to all walls including leaflet surfaces ($y+= 0.46$ at the peak flow) and predict the velocity gradients normal to the walls, prism layers utilized to allow the solver to resolve near wall flow accurately, Fig. 2c. In addition, uncertainty and error for this CFD simulation was calculated following ASME recommendations [47]. A fine-grid convergence index (GCIfine) and maximum discretization uncertainty in the area close to the leaflets were approximately 0.139% and 7%, respectively. These numerical uncertainties are comparable to previous studies [1]. The normalized velocity profile along a line located 7 mm downstream of the healthy valve (at the peak systole) is shown in Fig. 2b for a normal functioning valve. The validation of the velocity profiles in the current study was obtained by comparing the result with previous studies with similar geometries and flow conditions [39,40]. The root-mean-square (RMS) of the velocity differences between the current and the previous experimental study was 6.58% of the maximum velocity, suggesting agreement with measured values [39].

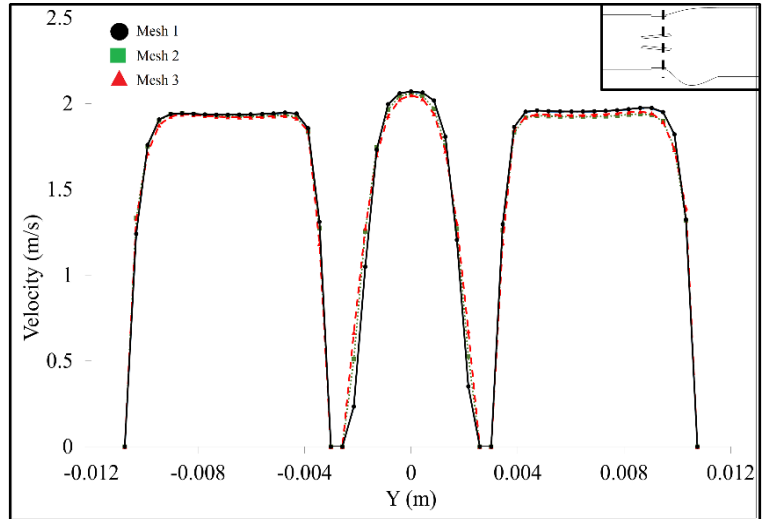

(a)

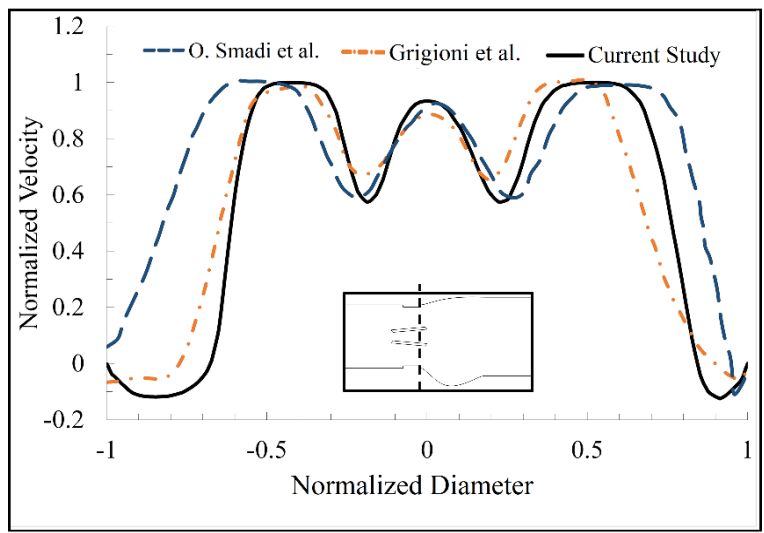

(b)

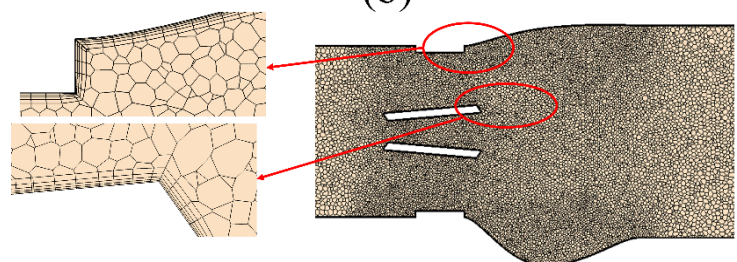

(c)

**Fig. 2** (a) Mesh-independent solution analysis with three different mesh configurations; (b) Normalized velocity distribution compared to previous CFD and experimental studies; (c) High Quality Polyhedral Mesh.

## 3. RESULTS AND DISCUSSION

Fig. 3 shows the velocity magnitude and fluctuation distributions through the valve with different levels of dysfunction at the peak systolic phase. The maximum velocity of 2.05 ms$^{-1}$ was computed for the healthy valve while it increased up to 4.49 ms$^{-1}$ as the leaflet dysfunction increased to 100%. The flow through the 0% dysfunctional valve was relatively uniform; however, the presence of the two valve leaflets lead to more shear layer generation than native valve flows. The valve with the one dysfunctional leaflet created high levels of disturbance in the flow. For the 50% valve dysfunction, development of the highest velocity (up to 3.3 m.s$^{-1}$) was observed through all three orifices. This resulted in large eddies between the healthy leaflet and the valve ring and small-sized vortices shed downstream of the dysfunctional leaflet. Consequently, the velocity fluctuations increased and intensified in the aortic sinuses. The forward jets can be associated with high shear stresses, located in shear layers past the leaflets and valve housing. This can lead to blood clotting and aortic stenosis. The maximum velocity increased significantly for the 100% leaflet dysfunction through the top orifice. In this case the bottom orifice was blocked followed by a large region of reversed flow at the leading edge of the valve leaflets. Large eddies were also present downstream of the aortic sinuses. High levels of fluctuations can also potentially give rise to flow-induced sound sources.

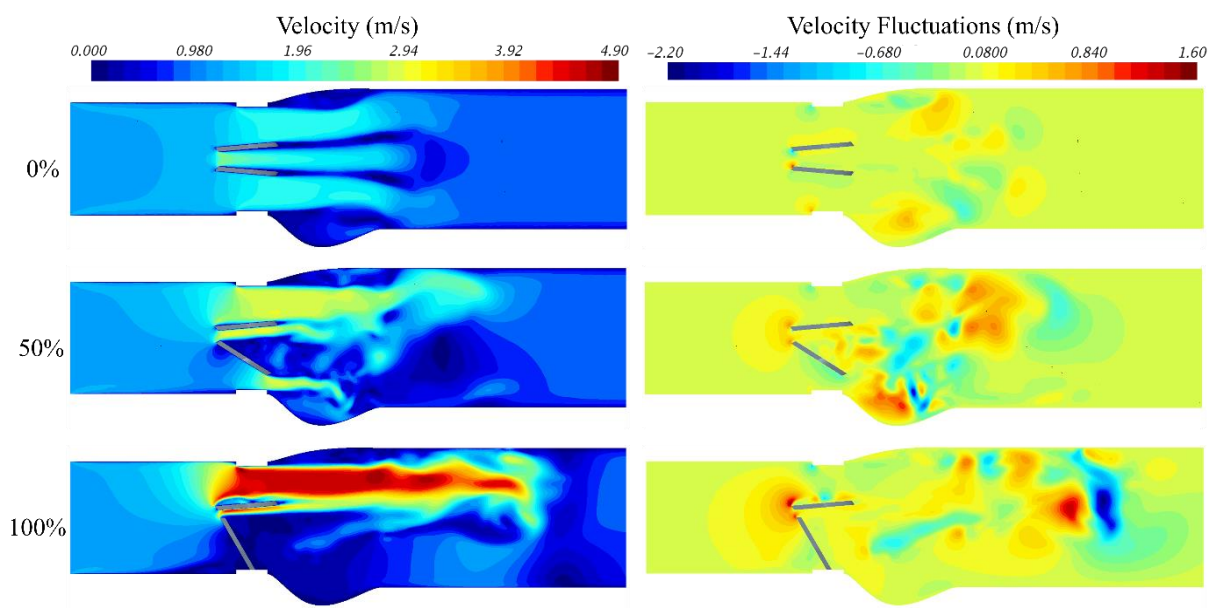

**Fig. 3** Velocity magnitude and fluctuations through the healthy and dysfunctional bileaflet mechanical heart valve

Fig. 4 shows the isosurface of vortical structures using lambda 2 criterion (also called "coherent turbulent structures" [48]) for the three valve dysfunction levels at the peak systolic phase. For the healthy valve (Fig. 4a), a region of vortices originated due to the blockage at the entrance of the valve. The regions with high levels of eddies were scarce, which demonstrates the desirable hemodynamic conditions of this BMHV. As also shown in Fig. 3, higher levels of vortical structures appear in the aortic sinuses region especially downstream of the dysfunctional leaflet for 50% valve dysfunction. At the sinus, the blood flow showed a strong recirculation region, which mixed with leaflet vortex wakes. This can lead to higher shear stresses on the aortic sinuses. For the 100% dysfunction, the high velocity through the top leaflet created a vortex shedding region downstream of the valve with low flow downstream of the bottom leaflet. Similar results can be found in [15].

Fig. 5 shows the distributions of wall shear stress (WSS) on aortic sinuses at different levels of dysfunctions at the peak systolic phase. The highest wall shear stresses occurred at the 50% dysfunction followed by 100%, and 0% dysfunctions, respectively. As shown in Fig. 3 and Fig. 4, the strength of the vortical structures significantly increased in addition to velocity fluctuations with dysfunction. For the 50% dysfunction, a high jet at the narrowing area of the bottom orifice caused concentrated vortices in the aortic sinuses region. These created a region of recirculation with high velocity fluctuations leading to the highest WSS about 165 N.m$^{-2}$. Conversely, at 100% dysfunctions, wall shear stresses on the sinus, especially downstream of the bottom leaflet, decreased (compared to 50% dysfunction) possibly due to flow obstruction by that leaflet. High wall shear stresses increase the potential risk of blood clotting and vascular diseases like aortic stenosis. The high stresses for the case of 50% dysfunction suggest that patients around that level of dysfunction may need closer monitoring.

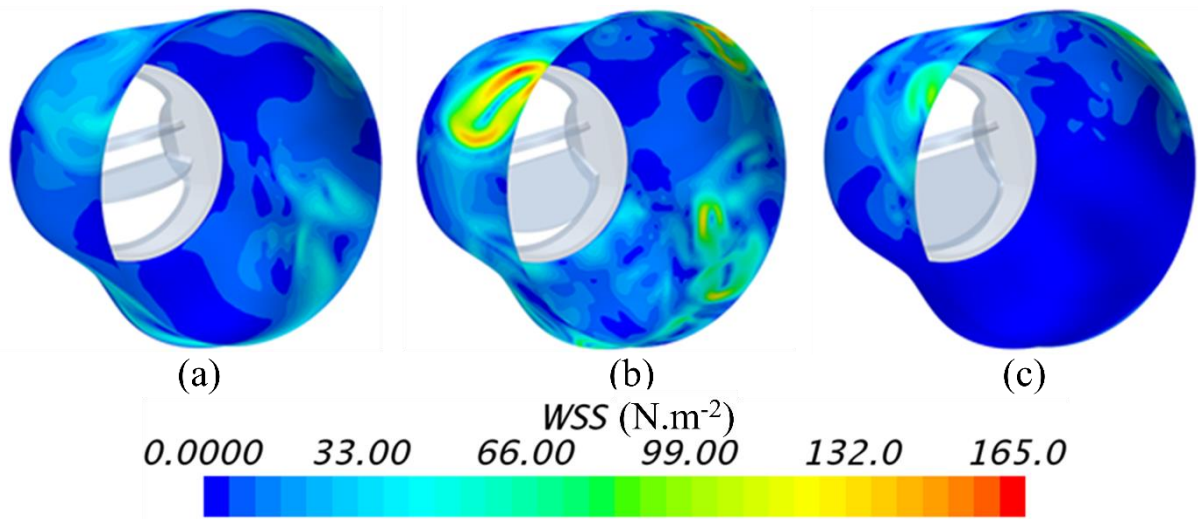

(a)  (b)  (c)

**WSS (N.m$^{-2}$)**

0.0000   33.00   66.00   99.00   132.0   165.0

**Fig. 5** Wall shear stress on the aortic sinuses for different levels of valve dysfunction at the peak systole.

Fig. 6 shows the negative velocity magnitudes in the flow direction (Fig. 6a, 6c, and 6e) and maximum turbulent shear stresses around the valves leaflets (Fig. 6b, 6d, and 6f) at the peak systole. A half-view of the volume was displayed to more clearly show the detailed vortical structures in the aortic sinuses. Several studies reported that the hemolysis (the breakage of a red blood cell membrane), can occur for turbulent shear stresses beyond the threshold of 400 N.m$^{-2}$ [32]. One potential region of high shear stresses is the recirculation area where velocity fluctuations exist along with low pressure gradients that may cause blood clotting [49]. The maximum turbulent shear stresses (Fig. 6 right side) were approximately 90, 420, and 800 N.m$^{-2}$ for 0%, 50%, and 100% dysfunction, respectively. On the other hand, the reversed flow (Fig. 6 left side) could attain velocity of 0.5, 1.6, and 2.1 m.s$^{-1}$. For the healthy valve, small region of recirculation occurred due to the step created by the valve ring and a sag caused by the shape of the sinuses. However, low velocity fluctuations in this region led to low levels of stresses and potential for blood damage. For the 50% dysfunction, the reverse flow in the aortic sinuses as well as downstream of the bottom leaflet increased. The results showed a strong flow recirculation in the sinuses which may explain why more highly damaged platelets were found in the aortic chamber [22]. This high flow recirculation zone almost filled the aortic sinuses. This region where coronary arteries originate is of particular importance as formation of clots at this location can lead to a heart attack. It was observed that the high stresses levels (which increase the potential risk for blood damage and platelet activation) were highest around the 50% dysfunction, while the maximum turbulent shear stress were close to the sinuses walls, which may further contribute to increased risk of blood clotting.

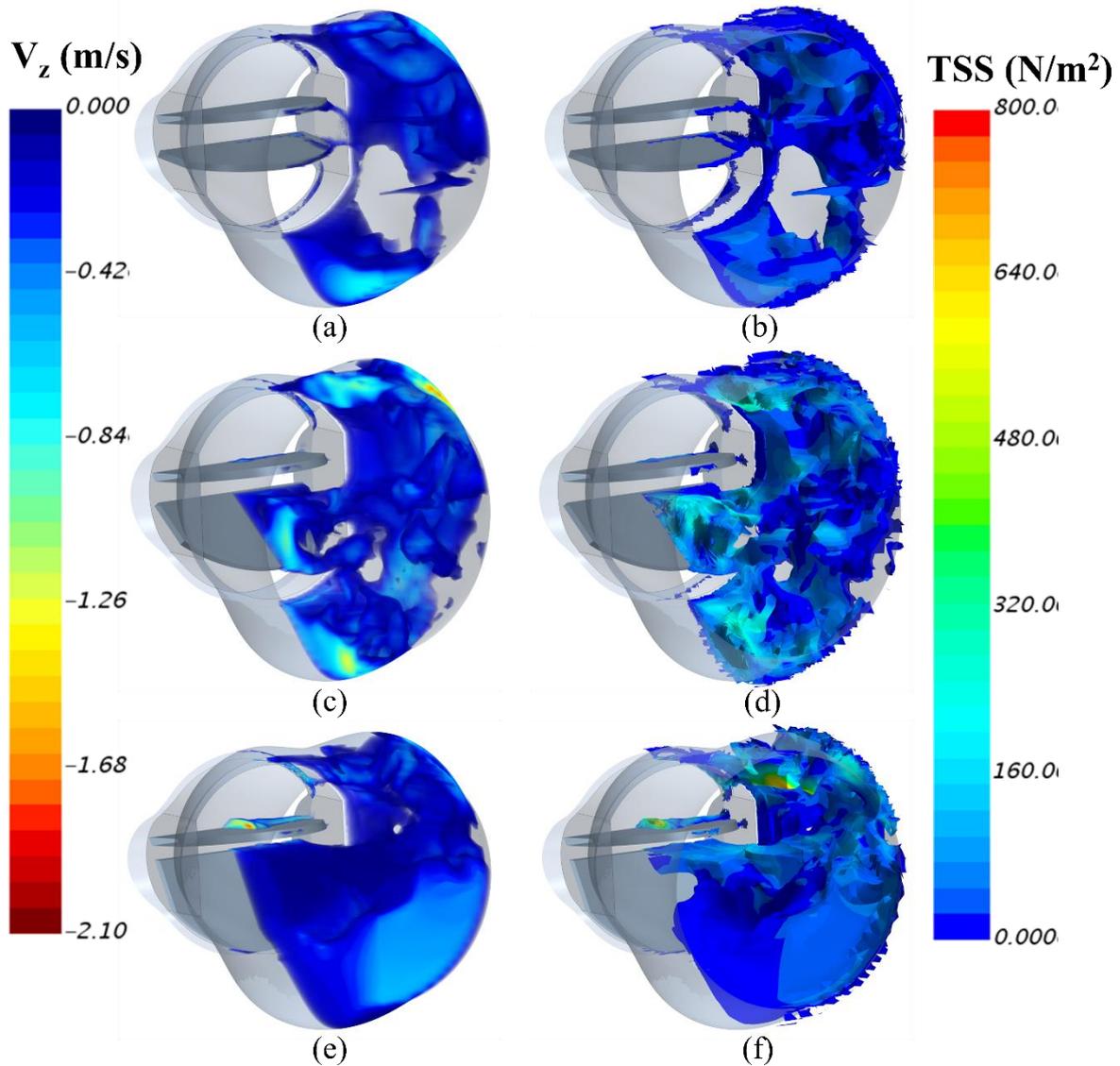

**Fig. 6** Flow recirculation and turbulent shear stresses downstream of the valve with 0% (a and b); 50% (c and d); and 100% (e and f) dysfunction.

## 6. CONCLUSIONS

In this study, blood flow through a bileaflet mechanical heart valve at different levels of leaflet dysfunction was analyzed. Results suggested increased velocities with dysfunction. For instance, as dysfunctionality increased the maximum velocity increased at the entrance of the aortic sinuses from 2.05 to 4.49 m/s (from 0% to 100% dysfunction) in the top orifices above the normal valve leaflet. Higher velocities and flow separation at the leaflet surfaces were accompanied by growing eddies demonstrated by increased vortical structures especially downstream of the valve. In addition, the maximum turbulent shear stresses reached up to 800 N.m$^{-2}$ exceeded the threshold values for elevated risk of hemolysis and platelet activation, which can lead to potential developing thrombosis, especially around the normal leaflet. In this study, the region in the aortic sinuses with large vortical structures and flow recirculation were also dealt with highest wall shear stresses for the 50% dysfunction level, suggesting that closer monitoring may be needed for patients around this level of dysfunction.